\documentclass{article}
\usepackage{spconf,amsmath,graphicx}
\usepackage{multirow}
\usepackage{float}
\usepackage{tikz}
\newcommand*\circled[1]{\tikz[baseline=(char.base)]{\node[shape=circle,draw,inner sep=2pt] (char) {#1};}}

\usepackage[colorlinks=true, urlcolor=blue, linkcolor=red]{hyperref}

\title{U-Net v2: Rethinking the Skip Connections of U-Net for Medical Image Segmentation\thanks{This research was supported in part by NIH NIBIB Grant R01-EB004640.}
}
\name{Yaopeng Peng$^{1}$ 
\qquad Milan Sonka$^{2}$
\qquad Danny Z. Chen$^{1}$}
  
\address{$^{1}$University of Notre Dame\qquad
\quad
$^{2}$University of Iowa
}
\begin{document}
%
\maketitle
\begin{abstract}
In this paper, we introduce U-Net v2, 
a new robust and efficient U-Net variant for medical image segmentation. It aims to augment the infusion of semantic information into low-level features while simultaneously refining high-level features with finer details. For an input image, we begin by extracting multi-level features with a deep neural network encoder. Next, we enhance the feature map of each level by infusing semantic information from higher-level features and integrating finer details from lower-level features through Hadamard product. Our novel skip connections empower features of all the levels with enriched semantic characteristics and intricate details. The improved features are subsequently transmitted to the decoder for further processing and segmentation. Our method can be seamlessly integrated into any Encoder-Decoder network. We evaluate our method on several public medical image segmentation datasets for skin lesion segmentation and polyp segmentation, and the experimental results 
demonstrate the segmentation accuracy of our new method over state-of-the-art methods, while preserving memory and computational efficiency. 
Code is available at: \href{https://github.com/yaoppeng/U-Net_v2}{https://github.com/yaoppeng/U-Net\_v2}. 

\end{abstract}
\section{Introduction}
\label{sec:intro}

With the advance of modern deep neural networks, significant progress has been made in 
semantic image segmentation. A typical paradigm for semantic image segmentation involves an Encoder-Decoder network with skip connections~\cite{long2015fully}. In this framework, the Encoder extracts hierarchical and abstract features from an input image, while the decoder takes the feature maps generated by the encoder and reconstructs a pixel-wise segmentation mask or map, assigning a class label to each pixel in the input image. A series of 
studies~\cite{zhao2017pyramid,liu2018path}
have been conducted to incorporate global information into the feature maps and enhance multi-scale features, resulting in substantial improvements in segmentation performance.

In the field of medical image analysis, accurate image segmentation plays a pivotal role in computer-aided diagnosis and analysis. U-Net~\cite{ronneberger2015u} was originally introduced for medical image segmentation, utilizing skip connections to connect the encoder and decoder stages at each level. The skip connections empower the decoder to access features from earlier encoder stages, hence preserving both high-level semantic information and fine-grained spatial details. This approach facilitates precise delineation of object boundaries and extraction of small structures in medical images. Additionally, a dense 
connection mechanism was applied 
to reduce dissimilarities between features in the encoders and decoders by concatenating features from all levels and all stages \cite{zhou2018unet++}. 
A mechanism was designed to enhance features by concatenating features of different scales from both higher and lower levels \cite{zhang2018mdunet}.

However, these connections in U-Net based models may not be sufficiently effective in integrating low-level and high-level features. For example, in ResNet~\cite{he2016deep}, a 
deep neural network was formed as an ensemble of multiple shallow networks, and an explicitly added residual connection illustrated that the network can struggle to learn  the identity map function, even when trained on a million-scale image 
dataset.

\begin{figure*}
\centering
\includegraphics[width=1.84\columnwidth]{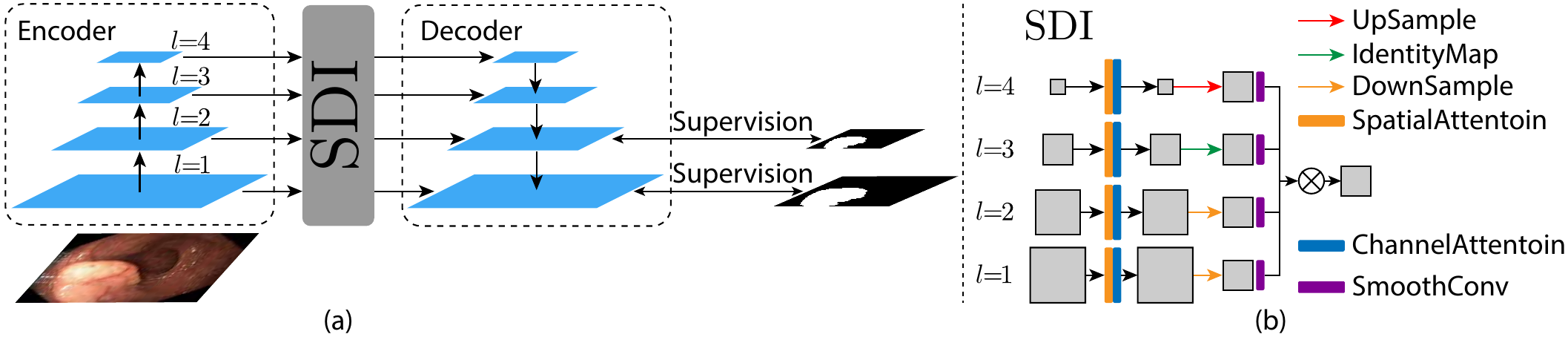}
\caption{(a) The overall architecture of our U-Net v2 model, which consists of an Encoder, the SDI (semantics and detail infusion) module, and a Decoder. (b) The architecture of the SDI module. For simplicity, we only show the refinement of the third level features ($l=3$). SmoothConv denotes a $3\times 3$ convolution for feature smoothing.  $\bigotimes$ denotes the Hadamard product.}
\label{fig:pipeline}
\end{figure*}

Regarding the features extracted by the encoders, the low-level features usually preserve more details but lack sufficient semantic information and may contain undesired noise. In contrast, the high-level features contain more semantic information but lack precise details (e.g., object boundaries) due to the significant resolution reduction. Simply fusing features through concatenation will heavily rely on the network's learning capacity, which is often proportional to the training dataset size. This is a challenging issue, especially in the context of medical imaging, which is commonly constrained by limited data. Such information fusion, accomplished by concatenating low-level and high-level features across multiple levels through dense connections, may limit the contribution of information from different levels and potentially introduce noise. On the other hand, despite the fact that the additional convolutions introduced do not significantly increase the number of parameters, GPU memory consumption will rise because all intermediate feature maps and the corresponding gradients must be stored for forward passes and backward gradient computations. This leads to an increase in both GPU memory usage and floating point operations (FLOPs).

In \cite{fan2020pranet}, reverse 
attention was 
utilized to explicitly establish connections among multi-scale features. In \cite{wei2021shallow}, ReLU activation was applied to higher-level features and the activated features were multiplied with lower-level features. Additionally, in \cite{zhang2021transfuse}, the authors proposed to extract features from CNN and Transformer models separately, combining the features from both the CNN and Transformer branches at multiple levels to enhance the feature maps. However, these approaches are complex, and their performance remains not very satisfactory, thus desiring further improvement.


In this paper, we present U-Net v2, a new U-Net based segmentation framework with straightforward and efficient skip connections. Our model first extracts multi-level feature maps using a CNN or Transformer encoder. Next, for a feature map at the $i$-th level, we explicitly infuse higher-level features (which contain more semantic information) and lower-level features (which capture finer details) through a simple Hadamard product operation, thereby enhancing both the semantics and details of $i$-th level features. Subsequently, the refined features are transmitted to the decoder for resolution reconstruction and segmentation. Our method can be seamlessly integrated into any Encoder-Decoder network. 

We evaluate our new method on two medical image segmentation tasks, Skin Lesion Segmentation and Polyp Segmentation, using publicly available datasets. The experimental results demonstrate that our U-Net v2 consistently outperforms state-of-the-art methods in these segmentation tasks while preserving FLOPs and GPU memory efficiency. 

\section{Method}

\label{sec:Methods}
\subsection{Overall Architecture}
The overall architecture of our U-Net v2 is shown in Fig.~\ref{fig:pipeline}(a). It comprises three main modules: the encoder, the SDI (Semantic and Detail Infusion) module, and the decoder.

Given an input image $I$, with $I \in R^{H\times W\times C}$, the encoder produces features in $M$ levels. We denote the $i$-th level features as $f_i^0$, $1 \leq i \leq M$. These collected features, $\{f_1^0, f_2^0, \ldots, f_M^0\}$, are then transmitted to the SDI 
module for further refinement.

\subsection{Semantics and Detail Infusion (SDI) Module}

With the hierarchical feature maps generated by the encoder, we first apply the spatial and channel attention mechanisms~\cite{woo2018cbam} to the features $f^0_i$ of each level $i$. This process enables the features to integrate both local spatial information and global channel information, as formulated below:
\begin{equation}
    f^1_i = \phi^c_i(\varphi^s_i(f^0_i)),
\end{equation}
where $f^1_i$ represents the processed feature map at the $i$-th level, and $\varphi^s_i$ and $\phi^c_i$ denote the parameters of spatial and channel attentions at the $i$-th level, respectively. Furthermore, we apply a $1\times 1$ convolution to reduce the channels of $f^1_i$ to $c$, where $c$ is a hyper-parameter. This resulted feature map is denoted as $f^2_i$, with $f^2_i\in R^{H_i\times W_i\times c}$, where $H_i$, $W_i$, and $c$ represent the width, height, and channels of $f^2_i$, respectively.

Next, we need to send the refined feature maps to the decoder. At each decoder level $i$, we use $f^2_i$ as the target reference. Then, we adjust the sizes of the feature maps at every $j$-th level to match the same resolution as $f^2_i$, formulated as:

\begin{equation}
f^3_{ij} =
\begin{cases}
\circled{D} (f^2_j, (H_i, W_i))& \text{if $j < i$ },\\[5pt]
\circled{I} (f^2_j) & \text{if $j=i$},\\[5pt]
\circled{U}(f^2_j, (H_i, W_i)) & \text{if $j>i$},
\end{cases}       
\end{equation}

\setlength{\tabcolsep}{3pt}
\begin{table}[t]
\centering
\begin{tabular}{c|cccc}
Dataset & Method & DSC (\%) & IoU (\%)  \\
\hline 

\multirow{5}{*}{ISIC 2017} & U-Net~\cite{ronneberger2015u} & 86.99& 76.98 \\

& TransFuse~\cite{zhang2021transfuse} & 88.40& 79.21 \\


& MALUNet~\cite{ruan2022malunet} & 88.13& 78.78  \\


& EGE-UNet~\cite{ruan2023ege}& 88.77 & 79.81 \\


\cline{2-4}
& U-Net v2 (ours)&\textbf{90.21} & \textbf{82.17}  \\


\hline 

\multirow{6}{*}{ISIC 2018}& U-Net~\cite{ronneberger2015u} & 87.55 & 77.86  \\

& UNet++~\cite{zhou2018unet++} & 87.83& 78.31 \\


& TransFuse~\cite{zhang2021transfuse}& 89.27&80.63 \\
& SANet~\cite{wei2021shallow} & 88.59& 79.52 \\



& EGE-UNet~\cite{ruan2023ege} & 89.04& 80.25 \\

\cline{2-4}
& U-Net v2 (ours) & \textbf{91.52} & \textbf{84.15} 


\end{tabular}

\caption{Experimental comparison with state-of-the-art methods on the two ISIC  datasets.}
\label{tab:isic}
\end{table}






\setlength{\tabcolsep}{3pt}
\begin{table}[h!]
\centering
\begin{tabular}{c|ccccc}
Datasets & Methods & DSC (\%) & IoU (\%) & MAE  \\

\hline 
\multirow{6}{*}{Kvasir-SEG} & U-Net~\cite{ronneberger2015u} &  81.8 & 74.6 & 0.055 \\

& UNet++~\cite{zhou2018unet++} & 82.1 & 74.3 & 0.048 \\

& PraNet~\cite{fan2020pranet} &  89.8 & 84.0 & 0.030  \\

& SANet~\cite{wei2021shallow}& 90.4 & 84.7 & 0.028  \\

& Polyp-PVT~\cite{dong2021polyp}& 91.7 & 86.4 & 0.023 \\
\cline{2-5}
& U-Net v2 (ours)&\textbf{92.8} & \textbf{88.0} & \textbf{0.019} \\

\hline 

\multirow{6}{*}{ClinicDB}& U-Net~\cite{ronneberger2015u} &  82.3 & 75.5 & 0.019 \\

& UNet++~\cite{zhou2018unet++} & 79.4 & 72.9 & 0.022 \\

& PraNet~\cite{fan2020pranet} & 89.9 & 84.9 & 0.009  \\

& SANet~\cite{wei2021shallow}& 91.6 & 85.9 & 0.012 \\

& Polyp-PVT~\cite{dong2021polyp}&93.7 & 88.9 & \textbf{0.006} \\
\cline{2-5}
& U-Net v2 (ours)& \textbf{94.4} & \textbf{89.6} & \textbf{0.006}  \\

\hline 

\multirow{6}{*}{ColonDB}& U-Net~\cite{ronneberger2015u} &  51.2 & 44.4 & 0.061 \\

& UNet++~\cite{zhou2018unet++} &48.3 & 41.0 & 0.064 \\

& PraNet~\cite{fan2020pranet} & 71.2 & 64.0 & 0.043   \\

& SANet~\cite{wei2021shallow}& 75.3 & 67.0 & 0.043 \\

& Polyp-PVT~\cite{dong2021polyp}& 80.8 & 72.7 & 0.031 \\
\cline{2-5}
& U-Net v2 (ours)&\textbf{81.2} & \textbf{73.1} & \textbf{0.030}  \\

\hline 

\multirow{6}{*}{ETIS}& U-Net~\cite{ronneberger2015u} &  39.8 & 33.5 & 0.036 \\

& UNet++~\cite{zhou2018unet++} &40.1 & 34.4 & 0.035 \\

& PraNet~\cite{fan2020pranet} & 62.8 & 56.7 & 0.031  \\

& SANet~\cite{wei2021shallow}& 75.0 & 65.4 & 0.015 \\

& Polyp-PVT~\cite{dong2021polyp}& 78.7 & \textbf{70.6} & \textbf{0.013} \\
\cline{2-5}
& U-Net v2 (ours)&\textbf{79.0} & {70.5} & \textbf{0.013}  \\
\hline 

\multirow{6}{*}{Endoscene}& U-Net~\cite{ronneberger2015u} &  71.0 & 62.7 & 0.022 \\

& UNet++~\cite{zhou2018unet++} &  70.7 & 62.4 & 0.018 \\

& PraNet~\cite{fan2020pranet} &  87.1 & 79.7 & 0.010  \\

& SANet~\cite{wei2021shallow}& 88.8 & 81.5 & 0.008 \\

& Polyp-PVT~\cite{dong2021polyp}& \textbf{90.0} & \textbf{83.3} & \textbf{0.007}\\
\cline{2-5}
& U-Net v2 (ours)&89.7 & 83.1 & \textbf{0.007}  \\
\end{tabular}

\caption{Experimental comparison with state-of-the art methods on the Polyp datasets.}
\label{tab:polyp}
\end{table}

where $\circled{D}$, $\circled{I}$, and $\circled{U}$ represent adaptive average pooling, identity mapping, and bilinearly interpolating $f^2_j$ to the resolution of $H_i\times W_i$, respectively, with $1\leq i, j  \leq M$.

Afterwards, a $3\times 3$ convolution is applied in order to smooth each resized feature map $f^3_{ij}$, formulated as:
\begin{eqnarray}
    f^4_{ij} = \theta_{ij}(f^3_{ij}),
\end{eqnarray}
where $\theta_{ij}$ represents the parameters of the smooth convolution, and $f^4_{ij}$ is the $j$-th smoothed feature map at the $i$-th level.

After resizing all the $i$-th level feature maps into the same resolution, we apply the element-wise Hadamard product to all the resized feature maps to enhance 
the $i$-th level features
with both more semantic information and finer details, as:
\begin{equation}
    f^5_i = H([f^4_{i1}, f^4_{i2}, \ldots, f^4_{iM}]),
\end{equation}
where $H(\cdot)$ denotes the Hadamard product (see Fig.~\ref{fig:pipeline}(b)). Afterwards, $f^5_i$ is dispatched to the $i$-th level decoder for further resolution reconstruction and segmentation.

\setlength{\tabcolsep}{3pt}
\begin{table}[h!]

\centering
\begin{tabular}{c|cccc}

Dataset & Method  & DSC (\%)& IoU (\%) \\
\hline 

\multirow{4}{*}{ISIC 2017}


& UNet++ (PVT)~\cite{zhou2018unet++} & 89.60$\pm$0.17& 81.16$\pm$ 0.07  \\

& U-Net v2 w/o SDI & 89.85$\pm$0.14 & 81.57$\pm$0.06 \\

& U-Net v2 w/o SC & 90.20$\pm$0.13 & 82.16$\pm$0.05 \\

\cline{2-4}

& U-Net v2 (ours) &\textbf{90.21$\pm$0.13} & \textbf{82.17$\pm$0.05} \\

\hline 



& UNet++ (PVT)~\cite{zhou2018unet++} & 78.0$\pm$4.3& 69.6$\pm$3.9  \\

\multirow{4}{*}{ColonDB} & U-Net v2 w/o SDI & 79.2$\pm$4.1& 71.5$\pm$3.7 \\

& U-Net v2 w/o SC & \textbf{81.3$\pm$3.7} & 72.8$\pm$4.0 \\


\cline{2-4}

& U-Net v2 (ours)& 81.2$\pm$3.9 & \textbf{73.1$\pm$4.4}

\end{tabular}

\caption{Ablation study on the ISIC 2017 and ColonDB datasets. SC denotes spatial and channel attentions.}
\label{tab:isic_ablation}
\end{table}


\section{Experiments}
\label{sec:Experiments}

\subsection{Datasets}
We evaluate our new U-Net v2 using the following datasets.
\vspace{0.04in}

\noindent
\textbf{ISIC Datasets}: Two datasets of skin lesion segmentation are used: ISIC 
2017~\cite{codella2019skin,berseth2017isic}, 
which comprises 2050 dermoscopy images, 
and ISIC 2018~\cite{codella2019skin}, 
which contains 2694 dermoscopy images. For fair comparison, we follow the train/test split strategy as outlined in~\cite{ruan2023ege}.

\noindent
\textbf{Polyp Segmentation Datasets}: Five datasets are used: Kvasir-SEG~\cite{jha2020kvasir}, ClinicDB~\cite{bernal2015wm}, ColonDB~\cite{tajbakhsh2015automated}, Endoscene \cite{vazquez2017benchmark}, and ETIS~\cite{silva2014toward}. For fair comparison, we use the train/test split strategy in \cite{fan2020pranet}. Specifically, 900 images from ClinicDB and 548 images from Kvasir-SEG are used as the training set, while the remaining images serve as the test set.


\subsection{Experimental Setup}
We conduct experiments on an NVIDIA P100 GPU with 
PyTorch. Our network 
is optimized using the Adam optimizer, with an initial learning rate = 0.001, $\beta_1$ = 0.9, and $\beta_2$ = 0.999. We employ a polynomial learning rate decay with a power of 0.9. The maximum number of training epochs is set to 300. The hyper-parameter $c$ is set to 32. As the approach 
in~\cite{ruan2023ege}, we report DSC (Dice Similarity Coefficient) and IoU (Intersection over Union) scores for the ISIC datasets. For the polyp datasets, we report DSC, IoU, and MAE (Mean Absolute Error) scores. Each experiment is run 5 times, and the averaged results are reported.
We use the Pyramid Vision Transformer (PVT)~\cite{wang2021pyramid} as the encoder for feature extraction.

\begin{figure}[h!]
\centering
\includegraphics[width=1\columnwidth]{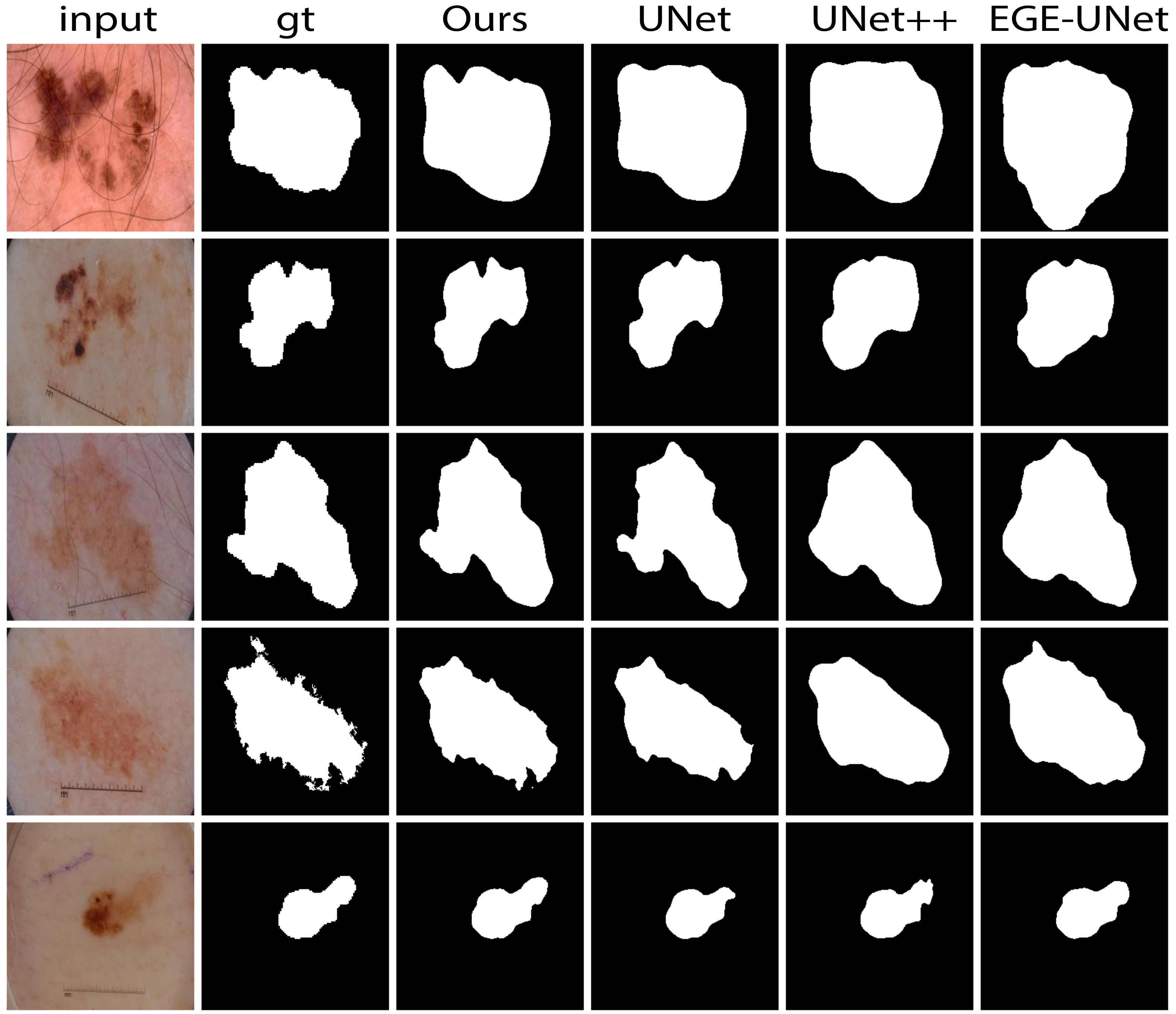}
\caption{Example segmentations from ISIC 2017 dataset. We use PVT as the encoder for U-Net and UNet++.
}
\label{fig:qualitative_result}
\end{figure}


\begin{table*}[h]
\centering
\begin{tabular}{c|c|c|c|c|c|c}

Model & DSC (ISIC 2017) & Input size& \# Params (M) & GPU memory usage (MB) & FLOPs (G) & FPS \\

\hline



U-Net (PVT) & 89.85 & (1, 3, 256, 256) & 28.15 & 478.82 & 8.433 & 39.678 \\

UNet++ (PVT) & 89.60 & (1, 3, 256, 256) & 29.87 & 607.31 & 19.121 & 34.431 \\

\hline

U-Net v2 (ours) & 90.21 & (1, 3, 256, 256) & 25.02 & 411.42 & 5.399 & 36.631

\end{tabular}
\caption{Comparison of computational complexity, GPU memory usage, and inference time, using an NVIDIA P100 GPU.}
\label{tab:complexity}
\end{table*}

\subsection{Results and Analysis}
Comparison results with state-of-the-art methods on the ISIC datasets are presented in Table~\ref{tab:isic}. As shown, our proposed U-Net v2 improves the DSC scores by 1.44\% and 2.48\%, and the IoU scores by 2.36\% and 3.90\% on the ISIC 2017 and ISIC 2018 datasets, respectively. These improvements demonstrate the effectiveness of our proposed method for infusing semantic information and finer details into each feature map.

Comparison results with state-of-the-art methods on the polyp segmentation datasets are presented in Table~\ref{tab:polyp}.
As shown, our proposed U-Net v2 outperforms Poly-PVT~\cite{dong2021polyp} on the Kavasir-SEG, ClinicDB, ColonDB, and ETIS datasets, with DSC score improvements of 1.1\%, 0.7\%, 0.4\%, and 0.3\%, respectively. This underscores the consistent effectiveness of our proposed method in infusing semantic information and finer details into feature maps at each level.

\subsection{Ablation Study}
We conduct ablation study using the ISIC 2017 and ColonDB datasets to examine the effectiveness of our U-Net v2, as reported in Table~\ref{tab:isic_ablation}. Specifically, we use the PVT~\cite{wang2021pyramid} model as the encoder for 
UNet++ \cite{zhou2018unet++}. Note that U-Net v2 is reverted to a vanilla U-Net with a PVT backbone when our SDI 
module is removed. SC denotes spatial and channel attentions within the SDI module. One can see from Table~\ref{tab:isic_ablation} that UNet++ exhibits a slight performance reduction compared to U-Net v2 without SDI (i.e., U-Net with the PVT encoder). This decrease may be attributed to the simple concatenation of multi-level features generated by dense connections, which could confuse the model and introduce noise. 
Table~\ref{tab:isic_ablation} demonstrates that the SDI module contributes the most to the overall performance, highlighting that our proposed skip connections (i.e., SDI) consistently yield performance improvements.

\subsection{Qualitative Results}
Some qualitative examples on the ISIC 2017 dataset are given in Fig.~\ref{fig:qualitative_result}, which demonstrate that our U-Net v2 is capable of incorporating semantic information and finer details into the feature maps at each level. Consequently, our segmentation model can capture finer details of object boundaries.

\subsection{Computation, GPU Memory, and Inference Time}

To examine the computational complexity, GPU memory usage, and inference time of our U-Net v2, we report the parameters, GPU memory usage, FLOPs, and FPS (frames per second) for our method, U-Net~\cite{ronneberger2015u}, and UNet++ \cite{zhou2018unet++}
in Table~\ref{tab:complexity}. 
The experiments use float32 as the data type, which results in 4B of memory usage per variable. The GPU memory usage records the size of the 
parameters and intermediate 
variables that are stored during the forward/backward pass. $(1, 3, 256, 256)$ represents the size of the input image. All the tests are conducted on an NVIDIA P100 GPU.

In Table~\ref{tab:complexity}, one can observe that UNet++ introduces more
parameters, and its GPU memory usage is larger due to the storage of intermediate variables (e.g., feature maps) during the dense forward process. Typically, such intermediate variables consume much more GPU memory than the parameters. Furthermore, the FLOPs and FPS of U-Net v2 are also superior to those of UNet++. The 
FPS reduction by our U-Net v2 compared to U-Net (PVT) is limited.

\section{Conclusions}
\label{sec:Conclusions}
A new U-Net variant was introduced, U-Net v2, which features a novel and straightforward design of skip connections for improved medical image segmentation. This design explicitly integrates semantic information from higher-level features and finer details from lower-level features into feature maps at each level produced by the encoder using a Hadamard product. Experiments conducted on Skin Lesion and Polyp Segmentation datasets validated the effectiveness of our U-Net v2. Complexity analysis suggested that U-Net v2 is also efficient in 
FLOPs and GPU memory usage.

\bibliographystyle{IEEEbib}
\bibliography{refs}

\begin{thebibliography}{10}

\bibitem{long2015fully}
Jonathan Long, Evan Shelhamer, and Trevor Darrell,
\newblock ``Fully convolutional networks for semantic segmentation,''
\newblock in {\em IEEE CVPR}, 2015, pp. 3431--3440.

\bibitem{zhao2017pyramid}
Hengshuang Zhao, Jianping Shi, Xiaojuan Qi, Xiaogang Wang, and Jiaya Jia,
\newblock ``Pyramid scene parsing network,''
\newblock in {\em CVPR}, 2017, pp. 2881--2890.

\bibitem{liu2018path}
Shu Liu, Lu~Qi, Haifang Qin, Jianping Shi, and Jiaya Jia,
\newblock ``Path aggregation network for instance segmentation,''
\newblock in {\em CVPR}, 2018, pp. 8759--8768.

\bibitem{ronneberger2015u}
Olaf Ronneberger, Philipp Fischer, and Thomas Brox,
\newblock ``{U-Net}: Convolutional networks for biomedical image
  segmentation,''
\newblock in {\em MICCAI, Proceedings, Part III}. Springer, 2015, pp. 234--241.

\bibitem{zhou2018unet++}
Zongwei Zhou, Md~Mahfuzur Rahman~Siddiquee, Nima Tajbakhsh, and Jianming Liang,
\newblock ``{UNet++}: A nested {U-Net} architecture for medical image
  segmentation,''
\newblock in {\em DLMIA 2018}. Springer, 2018, pp. 3--11.

\bibitem{zhang2018mdunet}
Jiawei Zhang, Yuzhen Jin, Jilan Xu, Xiaowei Xu, and Yanchun Zhang,
\newblock ``{MDU-Net}: Multi-scale densely connected {U-Net} for biomedical
  image segmentation,''
\newblock {\em arXiv preprint arXiv:1812.00352}, 2018.

\bibitem{he2016deep}
Kaiming He, Xiangyu Zhang, Shaoqing Ren, and Jian Sun,
\newblock ``Deep residual learning for image recognition,''
\newblock in {\em CVPR}, 2016, pp. 770--778.

\bibitem{fan2020pranet}
Deng-Ping Fan, Ge-Peng Ji, Tao Zhou, Geng Chen, Huazhu Fu, Jianbing Shen, and
  Ling Shao,
\newblock ``{PraNet}: Parallel reverse attention network for polyp
  segmentation,''
\newblock in {\em MICCAI}. Springer, 2020, pp. 263--273.

\bibitem{wei2021shallow}
Jun Wei, Yiwen Hu, Ruimao Zhang, Zhen Li, S~Kevin Zhou, and Shuguang Cui,
\newblock ``Shallow attention network for polyp segmentation,''
\newblock in {\em MICCAI, Proceedings, Part I 24}. Springer, 2021, pp.
  699--708.

\bibitem{zhang2021transfuse}
Yundong Zhang, Huiye Liu, and Qiang Hu,
\newblock ``{TransFuse}: Fusing {Transformers} and {CNNs} for medical image
  segmentation,''
\newblock in {\em MICCAI, Proceedings, Part I 24}. Springer, 2021, pp. 14--24.

\bibitem{woo2018cbam}
Sanghyun Woo, Jongchan Park, Joon-Young Lee, and In~So Kweon,
\newblock ``{CBAM}: Convolutional block attention module,''
\newblock in {\em ECCV}, 2018, pp. 3--19.

\bibitem{ruan2022malunet}
Jiacheng Ruan, Suncheng Xiang, Mingye Xie, Ting Liu, and Yuzhuo Fu,
\newblock ``{MALUNet}: A multi-attention and light-weight {UNet} for skin
  lesion segmentation,''
\newblock in {\em BIBM}. IEEE, 2022, pp. 1150--1156.

\bibitem{ruan2023ege}
Jiacheng Ruan, Mingye Xie, Jingsheng Gao, Ting Liu, and Yuzhuo Fu,
\newblock ``{EGE-UNet}: An efficient group enhanced {UNet} for skin lesion
  segmentation,''
\newblock {\em arXiv preprint arXiv:2307.08473}, 2023.

\bibitem{dong2021polyp}
Bo~Dong, Wenhai Wang, Deng-Ping Fan, Jinpeng Li, Huazhu Fu, and Ling Shao,
\newblock ``{Polyp-PVT}: Polyp segmentation with {Pyramid Vision
  Transformers},''
\newblock {\em arXiv preprint arXiv:2108.06932}, 2021.

\bibitem{codella2019skin}
Noel Codella, Veronica Rotemberg, Philipp Tschandl, M~Emre Celebi, Stephen
  Dusza, David Gutman, Brian Helba, Aadi Kalloo, Konstantinos Liopyris, Michael
  Marchetti, et~al.,
\newblock ``Skin lesion analysis toward melanoma detection 2018: A challenge
  hosted by the {International Skin Imaging Collaboration (ISIC)},''
\newblock {\em arXiv preprint arXiv:1902.03368}, 2019.

\bibitem{berseth2017isic}
Matt Berseth,
\newblock ``{ISIC} 2017-skin lesion analysis towards melanoma detection,''
\newblock {\em arXiv preprint arXiv:1703.00523}, 2017.

\bibitem{jha2020kvasir}
Debesh Jha, Pia~H Smedsrud, Michael~A Riegler, P{\aa}l Halvorsen, Thomas
  de~Lange, Dag Johansen, and H{\aa}vard~D Johansen,
\newblock ``{Kvasir-SEG}: A segmented polyp dataset,''
\newblock in {\em MMM, Part II 26}, 2020, pp. 451--462.

\bibitem{bernal2015wm}
Jorge Bernal, F~Javier S{\'a}nchez, Gloria Fern{\'a}ndez-Esparrach, Debora Gil,
  Cristina Rodr{\'\i}guez, and Fernando Vilari{\~n}o,
\newblock ``{WM-DOVA} maps for accurate polyp highlighting in colonoscopy:
  Validation vs. saliency maps from physicians,''
\newblock {\em CMIG}, vol. 43, pp. 99--111, 2015.

\bibitem{tajbakhsh2015automated}
Nima Tajbakhsh, Suryakanth~R Gurudu, and Jianming Liang,
\newblock ``Automated polyp detection in colonoscopy videos using shape and
  context information,''
\newblock {\em TMI}, vol. 35, no. 2, pp. 630--644, 2015.

\bibitem{vazquez2017benchmark}
David V{\'a}zquez, Jorge Bernal, F~Javier S{\'a}nchez, Gloria
  Fern{\'a}ndez-Esparrach, Antonio~M L{\'o}pez, Adriana Romero, Michal
  Drozdzal, Aaron Courville, et~al.,
\newblock ``A benchmark for endoluminal scene segmentation of colonoscopy
  images,''
\newblock {\em Journal of Healthcare Engineering}, vol. 2017, 2017.

\bibitem{silva2014toward}
Juan Silva, Aymeric Histace, Olivier Romain, Xavier Dray, and Bertrand Granado,
\newblock ``Toward embedded detection of polyps in {WCE} images for early
  diagnosis of colorectal cancer,''
\newblock {\em Journal of CARS}, vol. 9, pp. 283--293, 2014.

\bibitem{wang2021pyramid}
Wenhai Wang, Enze Xie, Xiang Li, Deng-Ping Fan, Kaitao Song, Ding Liang, Tong
  Lu, Ping Luo, and Ling Shao,
\newblock ``Pyramid {Vision Transformer}: A versatile backbone for dense
  prediction without convolutions,''
\newblock in {\em IEEE/CVF CVPR}, 2021, pp. 568--578.

\end{thebibliography}

\end{document}